\documentclass[preprint,12pt]{elsarticle}

\usepackage{amssymb}
\usepackage[T1]{fontenc}
\setcitestyle{super,open={},close={}}

\begin{document}

\begin{frontmatter}



\title{Electric Double Layer at the Metal-Oxide/Electrolyte Interface}


\author[inst1]{Lisanne Knijff}

\affiliation[inst1]{organization={Department of Chemistry-\AA{}ngstr\"{o}m Laboratory, Uppsala University},
            addressline={L\"{a}gerhyddsv\"{a}gen 1, BOX 538}, 
            city={Uppsala},
            postcode={75121}, 
            country={Sweden}}
\author[inst2]{Mei Jia}

\affiliation[inst2]{organization={Henan Key Laboratory of Biomolecular
    Recognition and Sensing, Henan Joint International Research
    Laboratory of Chemo/Biosensing and Early Diagnosis of Major
    Diseases, College of Chemistry and Chemical Engineering, Shangqiu Normal University}, addressline={Suiyang District}, city={Shangqiu}, postcode={476000}, country={China}
}%

\author[inst1]{Chao Zhang}
\ead{chao.zhang@kemi.uu.se}


\begin{abstract}
Metal-oxide surfaces act as both Br\o{}nsted acids and bases, which allows the exchange of protons with the electrolyte solution and generates either positive or negative proton charges depending on the environmental pH. These interfacial proton charges are then compensated by counter-ions from the electrolyte solution, which leads to the formation of the electric double layer (EDL). Because the EDL plays a crucial role in electrochemistry, geochemistry and colloid science, understanding the structure-property relationship of the EDL in metal-oxide systems from both experimental and theoretical approaches is necessary. This chapter focuses on the physical chemistry of the protonic double layer at the metal-oxide/electrolyte interface. In particular, determinations of the EDL capacitance and the double-layer potential from potentiometric titration experiments, electrochemical methods, surface-sensitive vibrational spectroscopy and X-ray photoelectron spectroscopy are summarized. This is followed by discussions from the atomistic modelling aspect of the EDL, with an emphasis on the density-functional theory-based molecular dynamics simulations. A conclusion and outlook for future works on this topic are also given.
\end{abstract}


\begin{highlights}
\item Introduces the physical chemistry of the protonic double layer at the metal-oxide/electrolyte interface.
\item Summarizes ways of determining the double-layer capacitance and the potential from experimental methods.
\item Discusses the atomistic modelling of the protonic double-layer, with an emphasis on the density-functional theory-based molecular dynamics simulations.
\end{highlights}


\end{frontmatter}



\section{Introduction}
What we are going to face in the 21st century is an increase of the worldwide energy consumption by about 100\% over the current level in the year of 2050~\cite{US_energy}. The corresponding yearly carbon emissions could reach about 75 gigatons by that time~\cite{OECD}. The explosion of energy needs and the experience of extreme climates in recent years suggest that we are indeed at a crossroad. Therefore, energy and environment are the two outstanding challenges that our society is facing.

One promising solution towards these challenges is to use metal oxide-based materials. Metal oxides are a class of inorganic materials that have various energy and environmental applications such as heterogeneous catalysts, fuel cells, lithium-ion batteries, supercapacitors, water treatment and antimicrobial application~\cite{Ardizzone:1996ca, Winter:2004ge, Simon:2008dc, Song.2018,Ren:2012go, Zheng:2018ie, Raghunath:2017hk}. Metal-oxide surfaces are terminated by O$^{2-}$ anions, the size of which is generally larger than that of the M$^{n+}$ cations, and this results in a lower coordination of surface metal cations with respect to the bulk and their accessibility to reactive environments. 

Most metal oxides are synthesized as nanostructures with different morphology and porosity, which leads to unique properties and reduced economical costs. For example, porous RuO$_2$ provides a very high specific capacitance of 1200 F g$^{-1}$ for supercapacitor applications due to the large surface area, the significant pore volume, and the high acid site concentration~\cite{Lokhande11}. On the other hand, the very properties that make metal oxide nanostructures attractive and indispensable in modern science and technology also cause an issue for the environment and human safety. There have been increasing concerns about the release of spent metal oxide nanostructures (e.g. ZnO nanoparticles as the most toxic one) into the aquatic environments via wastewater discharge~\cite{JOO201729}. 

In both the functioning and the degradation of metal oxide nanostructures, the aqueous interface plays a vital role~\cite{2017.Mu}. The pH condition in which the surface has no net proton charge is called the point of zero charge. Since working conditions (acid and alkaline solutions) for most metal oxides are not at the point of zero charge~\cite{Lyons:2017cg}, aqueous interfaces of metal oxide are highly electrified due to the acid-base chemistry and a protonic double layer is formed by attracting counter-ions from the electrolyte solution. The formation of these protonic double layers is a general phenomenon observed for all insulating main group oxides, such as silicates and aluminates as well as the conducting oxides formed by transition metal oxides. Compared to the metal/electrolyte interface, the surface charge density in these compact protonic double layers at the aqueous interface of oxides can be an order of magnitude larger~\cite{Sato1998}. Therefore, understanding the unique relationship between the protonic double-layer structure and the interfacial reactivity is a crucial step for designing metal oxide-based nanostructures. In the following, we will discuss the fundamentals of the protonic double layer from both experimental and theoretical aspects.

\section{Acid-base Chemistry and the Point of Zero Charge}
Acid-base reactions are a fundamental part of the metal-oxide chemistry, as they play an important role in the reactivity, the selectivity, and the formation of the protonic double layer. The idea of acid and base goes back to the 17th century where a base was seen as an anti-acid that can be made to react with an acid to form a salt. The modern understanding of the topic starts with Arrhenius (1884) who defined an acid as a compound that produces protons H$^+$ in aqueous solution and a base as a compound that produces hydroxyl ions OH$^-$ in aqueous solution. Later, the unique role of protons was highlighted by Br\o{}nsted (1923)~\cite{Bronsted23} and Lowry (1923)~\cite{Lowry23}, which laid the foundation for the acid-base chemistry.

In Br\o{}nsted-Lowry acid-base theory, an acid is defined as a proton-donor and a base is defined as a proton-acceptor, i.e.
\begin{equation}
    \label{eq:BL_definition}
    \mathrm{Acid} \rightleftharpoons \mathrm{Base} + \mathrm{H}^{+}
\end{equation}

This means that one can define acid-base pairs, where amphiprotic molecules belong to multiple acid-base pairs. It is also important to note that Eq.~\ref{eq:BL_definition} implies that there is always at least one ion-pair in an acid-base reaction, which entails the importance of electrostatic interactions in the determination of acidity.

To introduce the acidity or the p$K_a$, let us assume that the reaction that is being studied is:

\begin{equation}
\label{eq:pKa_reaction}
    \mathrm{HX} \mathrm{(aq)} + \mathrm{H_{2}O} \mathrm{(l)} \rightarrow \mathrm{H_{3}O}^{+} \mathrm{(aq)} + \mathrm{X}^{-} \mathrm{(aq)}
\end{equation}

Then, the corresponding reaction free energy $\Delta G$ is
\begin{equation}
    \label{eq:pKa_deltaG} 
    \Delta G = 2.30RT\left(\mathrm{p}K_a-\mathrm{p}K_{a,\mathrm{H}_3\mathrm{O}^+}\right)
\end{equation}

where \begin{equation}
\label{eq:pKa_def}
    \mathrm{p}K_{a} = -\mathrm{log}(K_a) = -\mathrm{log}\left(\frac{[ \mathrm{H}^{+}][ \mathrm{X}^{-}]}{[ \mathrm{HX}]}\right)
\end{equation}

In the Br\o{}nsted scale, $\mathrm{p}K_{a,\mathrm{H}_3\mathrm{O}^+} = -1.74$. Therefore, the significance of Eq.~\ref{eq:pKa_reaction} is that it recasts the problem of determining the p$K_a$ value into the free energy calculation of the proton transfer reaction, which is computationally feasible.~\cite{Cheng:2014jb}

Before moving on to surface acidity, it is worth mentioning a related quantity called proton affinity, which is defined as the enthalpy change for the following reaction:
\begin{equation}
\label{eq:pa_reaction}
    \mathrm{HX} \mathrm{(g)}  \rightarrow \mathrm{H}^{+} \mathrm{(g)} + \mathrm{X}^{-} \mathrm{(g)}
\end{equation}
When compared with Eq.~\ref{eq:pKa_reaction}, it is clear that the proton affinity is a gas-phase quantity whereas the p$K_a$ value depends on the solvent. 

Since metal oxides contain a wide variety of surface sites behaving as Br\o{}nsted acids and bases, reactions constantly occur when the surface comes into contact with an aqueous solution. The water molecules dissociate into protons and hydroxyl groups which strongly bind to the oxygen and metal atoms in the metal-oxide surface. Therefore, the p$K_{a}$ of a metal-oxide surface is commonly defined around the surface hydroxyls which are written here as MOH. Here the M is a metal atom, and the OH is the hydroxyl group bound to it. This hydroxyl group can act as an acid or a base by donating or adsorbing a proton respectively. The corresponding deprotonation reactions are as follows:

\begin{eqnarray}
\label{eq:pKa1}
\mathrm{MOH}_{2}^{+} \mathrm{(aq)} &\rightleftharpoons &
    \mathrm{MOH}  \mathrm{(aq)} +  \mathrm{H}^{+} \mathrm{(aq)},  {K}_{a1} \\
\label{eq:pKa2}
    \mathrm{MOH} \mathrm{(aq)} &\rightleftharpoons & \mathrm{MO}^{-} \mathrm{(aq)} + \mathrm{H}^{+} \mathrm{(aq)},  {K}_{a2}
\end{eqnarray}

Inspecting Eq.~\ref{eq:pKa1} and~\ref{eq:pKa2}, it is clear that the surface charge of the metal oxide depends on the concentration of H$^+$, or the pH of the aqueous solution.  

When the surface bears no net proton charge, the concentrations of $\mathrm{MOH}_{2}^{+}$ and $\mathrm{MO}^{-}$ will be equal. This means $K_{a1}$ and $K_{a2}$ would hold for the following expression:
\begin{equation}
\label{eq:pzc_conditon}
    K_{a1}K_{a2} = [\mathrm{H}^+]_{\mathrm{pzc}}^2
\end{equation}

\begin{figure} [h]
    \centering
    \includegraphics[width=0.8\linewidth]{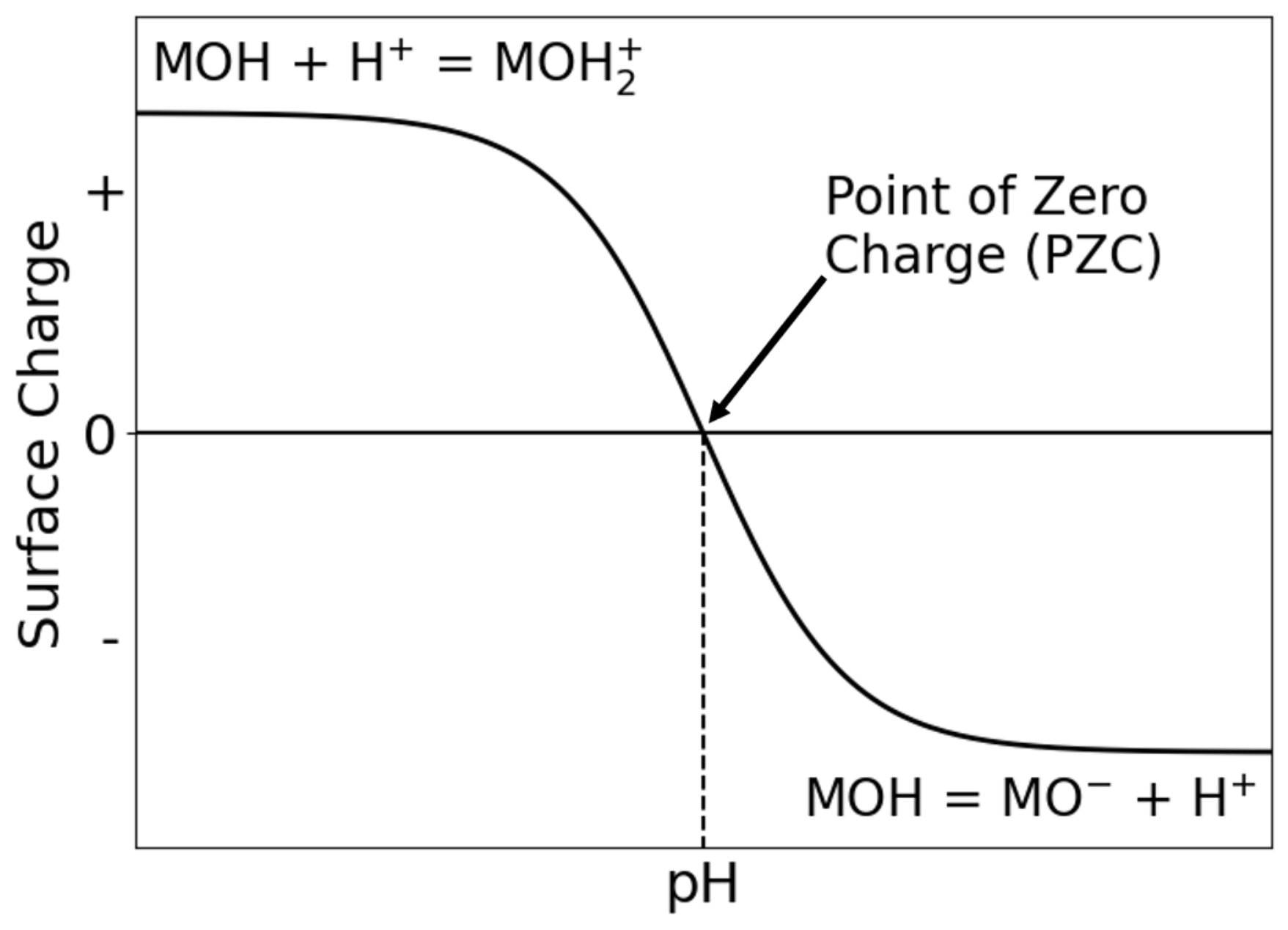}
    \caption{The point of zero charge at the metal-oxide/electrolyte interface.}
    \label{fig:pzc}
\end{figure}

Therefore, the corresponding pH for the neutral surface is
\begin{equation}
 \mathrm{pH_{pzc}} = -\log [\mathrm{H}^+]_\mathrm{pzc} = \frac{\mathrm{p}K_\mathrm{a1}+\mathrm{p}K_\mathrm{a2}}{2}
\end{equation}
This is the so-called the point of zero charge (PZC). Fig.~\ref{fig:pzc} presents a schematic diagram illustrating the manner in which the surface charge changes with pH. 

For metal-oxides, there are several physical parameters that determine the PZC. Such parameters, for example, are the cation electronegativity, the cation ionic radius and its formal charge. The oxygen partial charge, and the surface site coordination also play an important role. However, their direct link is hard to describe using theory and thus mainly relies on empirical models. \cite{noguera1996physics}

In the literature, the PZC and the isoelectric point (IEP) are sometimes used interchangeably. However, they are quite different quantities in nature because the referred charges are different. The PZC refers to the surface charge, while the IEP refers to the diffuse charge in the double layer.~\cite{lyklema2005fundamentals} This leads us to the discussion of the double-layer structure at the metal-oxide/electrolyte interface as outlined in the next section.

\section{The Formation of Electric Double Layer}

The electric double layer (EDL) manifests itself at both the metal/electrolyte and the metal-oxide/electrolyte interfaces. The EDL at the metal/electrolyte interface is a central topic in electrochemistry~\cite{schmickler2010interfacial}, while the EDL at the metal-oxide/electrolyte interface plays a crucial role in geochemistry and colloid science~\cite{bunker_aqueous_2020}. 

\begin{figure}[h]
    \centering
    \includegraphics[width=0.9\linewidth]{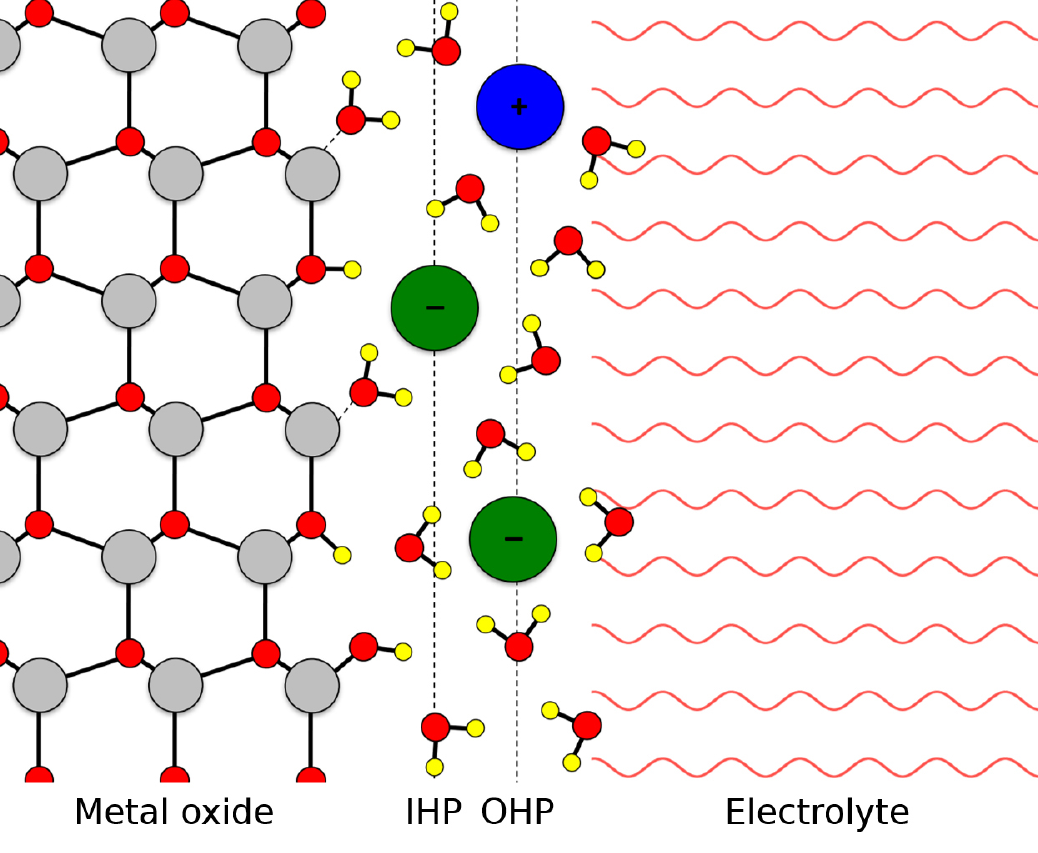}
    \caption{Model for the double layer region at the metal-oxide/electrolyte interface IHP, inner Helmholtz plane; OHP, and outer Helmholtz plane.}
    \label{fig:EDL}
\end{figure}

Similar to the electrified metal surface, adsorption of ions at the metal-oxide/electrolyte interface leads to the formation of the EDL, which is schematically represented in Fig.~\ref{fig:EDL}. The first layer of the double layer consists of a layer of solvent molecules and ions that are directly adsorbed to the electrified surface. The ions that are part of this so-called inner Helmholtz layer, no longer have their full solvation shell intact and the average position of those ions corresponds to the inner Helmholtz plane (IHP). Then, ions that still have their full solvation shell are not able to get as close to the electrified surface. The average position of those ions is called the outer Helmholtz plane (OHP). Finally, the long-range electrostatic interactions of the electrolyte with the electrified interface lead to the formation of a whole concentration profile of solvated ions over a large distance. This is called the diffuse layer or the Gouy-Chapman layer, and reaches from the outer Helmholtz plane all the way to the bulk of the electrolyte.~\cite{memming2015semiconductor} Despite of these great similarities between the EDL at the metal/electrolyte interface and that at the metal-oxide/electrolyte interface, there are also significant differences in several aspects.  

The first difference between the EDL at the metal-oxide/electrolyte interface as compared to that at the metal/electrolyte interface is the charging mechanism. The excess charge at the metal surface is due to the rearrangement of the electronic density under an external perturbation (e.g. a voltage source). In contrast, the excess charge at the metal-oxide surface is generated by acid-base chemistry, as discussed in the previous section. Apart from these acid/base interactions, other reactions occur as well. Since it is possible for some of the metal ions to dissolve in the electrolyte solution because of the pH condition, these can then react with deprotonated surface hydroxyls to form surface complexes instead. Moreover, these reactions happen exactly within the Helmholtz layer, which would alter the Helmholtz capacitance considerably. Nevertheless, it is worth noting that the distinction between the electronic charge versus the proton charge may not be clear-cut. For conducting oxides, such as RuO$_2$, both electronic charge and proton charge can exist simultaneously. In addition, most metal surfaces will be oxidized at a higher potential. Therefore, a combination of both charging mechanisms is more of a rule than an exception.

The second difference comes from the addition of the space-charge (SC) layer in metal-oxides. The space-charge layer is a result of the fact that metal-oxides have a much lower electronic conductivity, and less free electrons in general. Because of that, charging at the surface can no longer instantaneously be compensated, and instead leads to the depletion or accumulation of electrons at the metal-oxide surface. The layer where this is the case is referred to as the SC layer, which can have a thickness of 10-100 nm.~\cite{memming2015semiconductor}

\begin{figure}[h]
    \centering
    \includegraphics[width=0.8\linewidth]{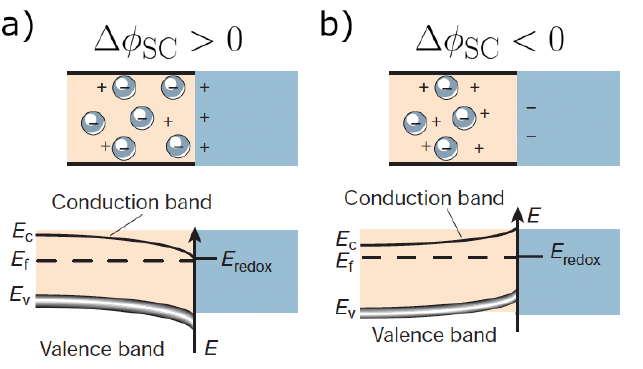}
    \caption{The \textit{n}-type semiconductor/electrolyte interface. a) Accumulation layer causes downward bending of the conduction and valence bands.  b) Depletion layer causes upward bending of the conduction and valence bands. Adapted from Ref.~\cite{2001.Graetzel} with permission (Copyright 2019 the Springer Nature Group)}
    \label{fig:potential_charge_interface}
\end{figure}

The electrostatic potential difference at the interface with respect to that in the bulk of the semiconductor is $\Delta\phi_\mathrm{SC}$. In the case of an $n$-type semiconducting oxide, the concentration of electrons in the conduction band is enhanced when $\Delta\phi_\mathrm{SC}$ is positive. Vice versa, the concentration of electrons at the surface is reduced (and that of holes increased)  when $\Delta\phi_\mathrm{SC}$ is negative, as shown in Fig.~\ref{fig:potential_charge_interface}. The situations of $p$-type semiconducting oxides are simply the opposite. Note that the band bends downwards ($\Delta\phi_\mathrm{SC} > 0$) and upwards ($\Delta\phi_\mathrm{SC}<0$) depending on the sign of $\Delta\phi_\mathrm{SC}$, and that the potential at which $\Delta\phi_\mathrm{SC} \approx 0$ is called the flat-band potential. 

Taking these factors into consideration, this means that the capacitance of a metal-oxide/electrolyte interface consists of three components, i.e. the capacitance of the space-charge layer $C_\mathrm{SC}$, the capacitance of the Helmholtz layer $C_\mathrm{H}$ and the capacitance of the Gouy-Chapman layer $C_\mathrm{GC}$, as follows:

\begin{equation}
    \frac{1}{C_{\mathrm{EDL}}} = \frac{1}{C_{\mathrm{SC}}} + \frac{1}{C_{\mathrm{H}}} + \frac{1}{C_{\mathrm{GC}}}
    \label{eq:MO_capacitance}
\end{equation}

The capacitance of the space-charge layer is similar to that of the Gouy-Chapman layer in the electrolyte. They are analogous, where the electrons in the space-charge layer can be seen as the ions in the Gouy-Chapman layer. Therefore, the expression for the capacitance can be obtained in a similar fashion by solving the one-dimensional Poisson-Boltzmann equation. This results in the so-called Mott-Schottky relation:

\begin{equation}
    \label{eq:C_sc}
    \frac{1}{C^2_\mathrm{SC}} = \left(\frac{2L_{D}}{\epsilon\epsilon_0}\right)^2\left(\frac{e\Delta\phi_\mathrm{SC}}{kT}-1\right)
\end{equation}

where $\epsilon$ is the dielectric constant, $\epsilon_0$ is the vacuum permittivity, $k$ is the Boltzmann factor and $T$ is the temperature. $L_D$ therein is the Debye length at a given concentration of the bulk electron concentration $n_0$ with the following expression

\begin{equation}
\label{eq:debye_length_SC}
    L_D=\left(\frac{\epsilon\epsilon_0kT}{2n_0e^2}\right)^{1/2}
\end{equation}

Compared to the space-charge layer and the Gouy-Chapman layer, the Helmholtz layer involves both short-range and long-range interactions. This makes it an indispensable factor that affects the chemical reactivity at the electrified metal-oxide/electrolyte interface. Moreover, $C_\mathrm{H}$ is also the dominating term in Eq.~\ref{eq:MO_capacitance} at a high salt concentration and the flat potential condition ($\Delta\phi_\mathrm{SC} = kT/e$). Therefore, we will see how $C_\mathrm{H}$ can be determined from both experimental and computational methods in the following sections.

\section{Double-layer Capacitance and Potential from Experiments}

The most used technique for determining the surface charge and the double layer capacitance at metal-oxide/electrolyte interfaces is the potentiometric titration method. 

In potentiometric titration experiments, the surface charge density $\sigma_0$ is determined from the surface concentrations $\Gamma_\mathrm{H^+}$ and $\Gamma_\mathrm{OH^-}$ of H$^+$ and OH$^-$ species respectively, as 

\begin{equation}
    \label{eq:charge_titration}
    \sigma_0 = F\left(\Gamma_{\mathrm{H}^+}-\Gamma_{\mathrm{OH}^-}\right)
\end{equation}

\begin{figure}[h]
    \centering
    \includegraphics[width=1.0\linewidth]{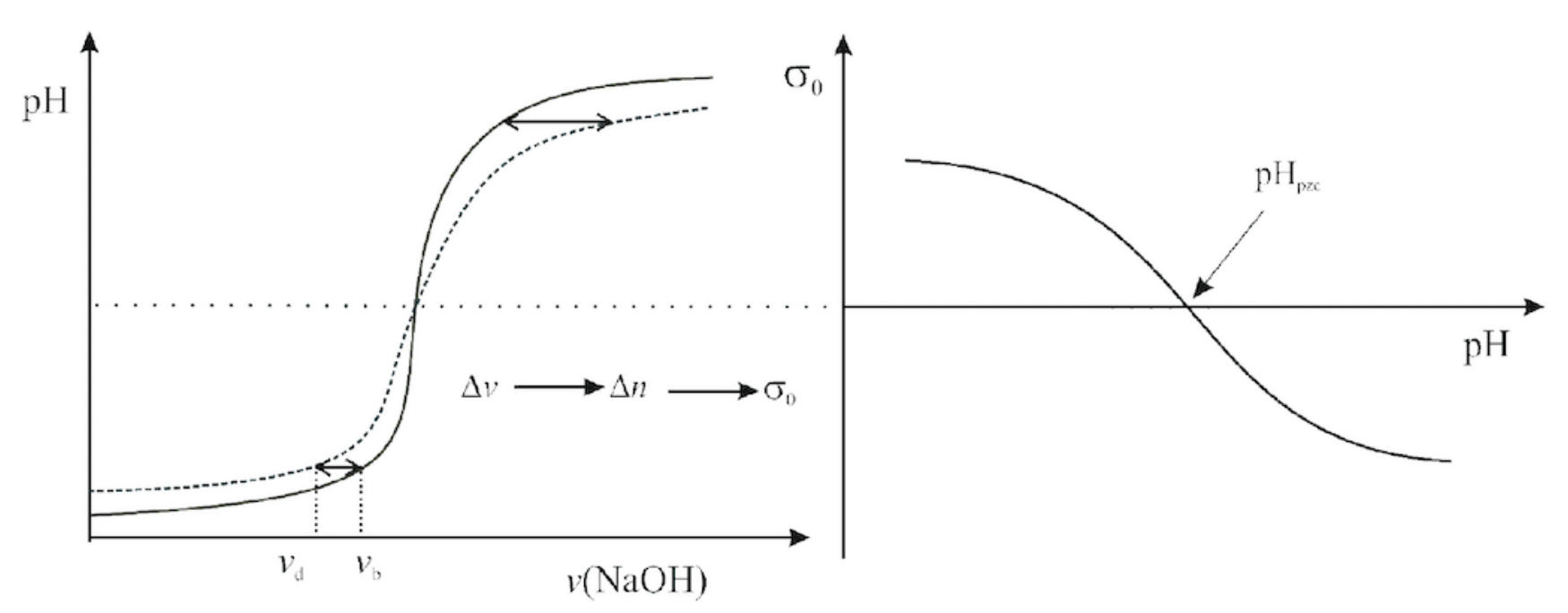}
    \caption{Schematic representation of the potentiometric titrations (left) of a colloid suspension (dashed line) and a blank titration without colloid particles (full line) and the resulting proton related surface charge density (right) as a function of pH. Adapted from Ref.~\cite{2012luetzenkirchen} under the terms of CC-BY licence (Copyright 2012 Authors).}
    \label{fig:titration_exp}
\end{figure}

The surface concentration $\Gamma$ therein is measured from the volume change $\Delta v = v_\mathrm{b} - v_\mathrm{d}$ of the titrant added to the solution before and after the dispersion of metal oxide nanoparticles at a constant pH condition, as illustrated in Fig.~\ref{fig:titration_exp}. Note that the specific surface area is needed in the conversion from $\Delta v$ to $\Gamma$, which is either assumed a prior or measured by separate experiments~\cite{2012luetzenkirchen}.

Subsequently, the capacitance can be obtained via the following formula~\cite{Berube:1968im}:
\begin{equation}
    C = \frac{d\sigma_0}{d\psi} = \frac{Fd(\Gamma_{\mathrm{H}^+}-\Gamma_{\mathrm{OH}^-})}{2.30RTd\mathrm{pH}}
\end{equation}

For semi-conducting oxides, impedance spectroscopy, which measures the resistive response to a time-varying electric field~\cite{lvovich_impedance_2015}, can also be applied. The impedance, as the ratio between the input voltage and the output current, has a real and imaginary components, which are directly related to the frequency-dependent conductivity and capacitance. These components are commonly represented in a complex plane as the so-called Nyquist plot. Then, equivalent circuit models are used to separate the contributions from the double layer and the space-charge layer to the overall capacitance~\cite{Tomkiewicz:1979ifa}.

For conducting oxides, such as RuO$_2$, cyclic voltammetry can also be used, in which the current response is recorded with respect to a linear potential sweep~\cite{faulkner2002electrochemical}. Since the potential increases (or decreases) linearly with time in cyclic voltammetry, the corresponding surface charge can be calculated by integrating the current. Thus, the $I-V$ curve can be readily converted to $dQ/dV-V$ curve, which shows the potential-dependence of the capacitance~\cite{2006.Sugimoto}. However, it should be pointed out that the capacitance obtained in this way is usually attributed as the pseudo-capacitance, which is a result of surface redox reactions. 

In addition to the titration and electrochemical methods mentioned above, surface-sensitive vibrational spectroscopy and X-ray photoelectron spectroscopy can also be used to probe the double-layer potential in the Helmholtz layer. They distinguish themselves from electrokinetic techniques that measure the zeta potential (the diffuse-layer potential). 

In the second harmonic generation (SHG)/sum-frequency generation (SFG) method, the polarization response is zero for any centro-symmetric environments, which makes the technique interface-selective. Then, the polarization at electrified metal-oxide/electrolyte interfaces contains the contributions from both the second-order susceptibility $\chi^{(2)}$ and the third-order susceptibility $\chi^{(3)}$~\cite{Ong:1992ca, 2020.Bischoff}. 

\begin{equation}
    \label{eq:SHG_EDL}
    P_{2\omega} = \chi^{(2)}E_\omega E_\omega + \chi^{(3)}\Delta\phi_\mathrm{EDL}E_\omega E_\omega
\end{equation}

The second-order susceptibility $\chi^{(2)}$ couples to the interfacial water and the surface hydroxyl groups in the frequency of O-H stretching, while the third-order susceptibility $\chi^{(3)}$ is related to the double-layer potential $\Delta\Phi_\mathrm{EDL}$. Therefore, $\Delta\phi_\mathrm{EDL}$ can be measured at a given pH condition from SHG/SFG experiments and the surface p$K_a$ can be determined accordingly.

In X-ray photoelectron spectroscopy, high-energy X-rays are used to probe the sample, which can cause electrons to be ejected from the probed species. The kinetic energy of those electrons can then be used to determine the binding energy (BE) of the electron in the same material. Therefore, the change in BE of metal-oxide nanoparticles at a given pH with respect to the PZC gives access to the double-layer potential~\cite{Brown:2016bo, Brown:2016kwa}.

\begin{equation}
    \label{eq:BE_edl}
    \mathrm{BE} - \mathrm{BE}_\mathrm{PZC} = e\Delta\phi_\mathrm{EDL}
\end{equation}

Subtracting the zeta potential due to the diffuse layer and incorporating the surface charge density measured from the potentiometric titration experiment, this provides another means to estimate the double-layer capacitance of the Helmholtz layer.

\section{Atomistic Modelling of the Protonic Double Layer}
Modelling of the protonic double layer at the metal-oxide/electrolyte interface started around the 1960s by relating the PZC with the formal ionic charge of the cation, and its ionic radius~\cite{1965.Parks}. There, it was found that basic oxides consist of cations with a low formal charge. High formal charges on the other hand result in acidic oxides. This emphasizes again the important role of the electrostatic interaction in the metal-oxide systems and triggered the development of surface complexation models (SCMs) from the 1970s~\cite{1968.Schindler, Stumm70, DAVIS:1978fn, 1980.Westall}.  Among those, the multisite complexation model (MUSIC)~\cite{Hiemstra:1989vh} and its charge-distribution variant (CD-MUSIC)~\cite{hiemstra1996surface} are most notable ones. In the MUSIC model, the phenomenological surface sites are taken into account to describe the adsorption and desorption of protons following the 2-pK model shown in Eqs.~\ref{eq:pKa1} and~\ref{eq:pKa2}, and the surface charges upon protonation/deprotonation are determined using the Pauling bond valence model (i.e. the charge of the metal cation is divided by the number of coordinating groups).  Later, it was extended in the CD-MUSIC model, where the charge of the inner-sphere complexes in the double layer is distributed over the surface plane and the next electrostatic plane in order to match the macroscopic adsorption data. In these models, the double layer is treated as a series of parallel-plate capacitors for adjusting the surface concentration of ions. 

Despite that the MUSIC and the CD-MUSIC model have been successfully applied to study different oxide materials such as silica, hematite, cerium oxide, rutile and anatase among others~\cite{2001brown}, it is worth noting that the binding constant (for proton adsorption/desorption), the Helmhotlz capacitance and the p$K_a$ value are fitting parameters to reproduce the potentiometric titration data. This raises questions about the physical soundness of surface complexation modelling~\cite{2006.Bickmore}.  Moreover, SCMs are purely thermodynamic models, the surface heterogeneity and the temporal evolution of the double layer are simply out of reach. This calls for atomistic modelling, which can provide predictive structural, dynamical and energetic information.

In particular, density-functional theory-based molecular dynamics (DFTMD) simulations~\cite{Car:1985ix,marx09} are in principle  well-suited for modelling the protonic double layer, where the distinction between reactive solutes and solvent has all but disappeared. In DFTMD, the force on the atom is computed ``on-the-fly'' by solving the electronic Schr\"odinger equation during the time evolution of the system. Therefore, bond forming/breaking, electronic polarization and charge transfer effects are all included. This is in contrast with MD simulations based on classical force fields~\cite{Predota:2004cy,Parez:2014dx}, in which the parameters used in predefined functional forms to describe the bonded and non-bonded interactions are obtained by fitting either structural and spectroscopic data of simple hydrated compounds or quantum chemical calculations~\cite{Bandura:2003fq, cygan2004molecular, 2013.Heinz}. 

While DFTMD can be highly accurate, it is also computationally very demanding. This makes DFTMD simulations of the electrified solid-liquid interface seemingly impractical.  However, endeavours have been attempted to address this challenging problem~\cite{cheng2014electric, ulman2019understanding, zhang2019coupling, jia2021origin}, and breakthroughs have recently been made by the development of finite-field DFTMD~\cite{zhang2020modelling}.

Since periodic boundary conditions are applied to the simulation cell (also known as the supercell), there are naturally two choices for setting up the protonic double layer in a DFTMD simulation. In the symmetric setup, two sides of the metal-oxide slab bear the same amount and the same type of proton charge and there is no supercell dipole in the overall system. Instead, in the asymmetric setup, two sides of the metal-oxide slab take the same amount but opposite types of proton charge and this leads to a net polarization in the system. The symmetric setup for the DFTMD modelling of the protonic double layer has been applied to systems of TiO$_2$(110)/NaCl solution~\cite{cheng2014electric} and hematite/NaF solution~\cite{ulman2019understanding}. Cheng and Sprik showed that the double-layer charging has a significant effect on the structure of the rutile TiO$_{2}$(110)/water interface using DFTMD, as the water molecule are forced to align their dipoles with the electric field. Values found for the capacitance are lower by a factor of 3 than the experimental value of 100~$\mu$F cm$^{-2}$, although there is also a large uncertainty in the experimental estimation~\cite{cheng2014electric}. Similarly, Ulman et al. adopted the symmetric setup in their DFTMD simulations of the hematite/NaF electrolyte interface and found that for the hydroxylated surface the value of the Helmholtz capacitance is about 40.3 $\mu$F cm$^{-2}$ while it changes to 51 $\mu$F cm$^{-2}$ for the FeO$_{3}$Fe termination~\cite{ulman2019understanding}. 

\begin{figure}[ht]
    \centering
    \includegraphics[width=1.0\linewidth]{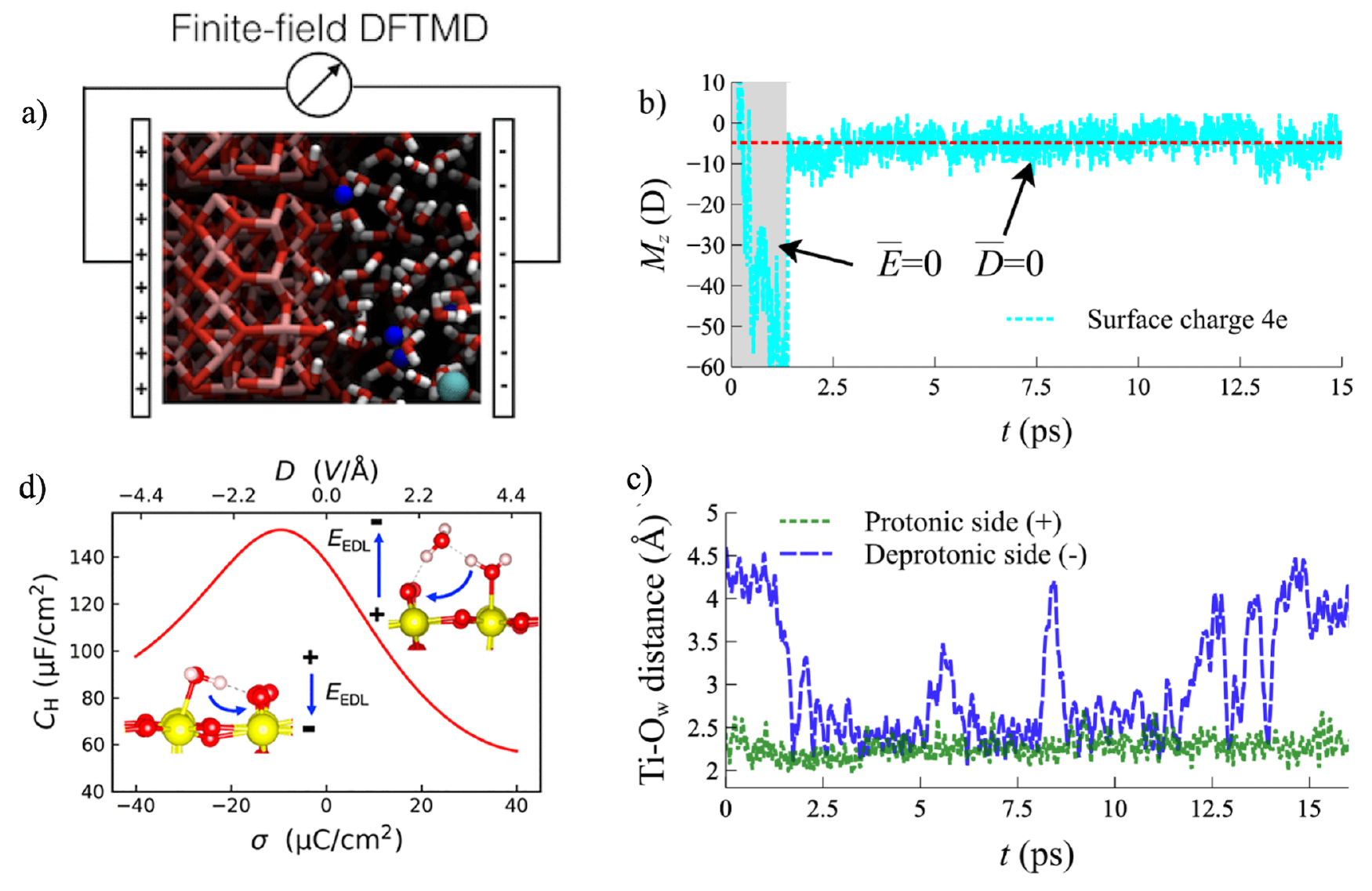}
    \caption{a) An illustration of the finite-field DFTMD applied to the rutile TiO$_2$(110)/NaCl electrolyte interface; b) The effect of the electric boundary condition on the time evolution of the supercell polarization; c) The water dynamics at the electrified interface of the rutile TiO$_2$ (110); d) The differential Helmholtz capacitance resolved from the finite-field DFTMD simulation. Adapted from Ref.~\cite{zhang2019coupling} with permission (Copyright 2019 American Chemical Society) and Ref.~\cite{jia2021origin} under the terms of CC-BY licence (Copyright 2021 Authors).}
    \label{fig:atomistic_EDL}
\end{figure}

Compared to the symmetric setup of the supercell, the asymmetric setup has the merit that the chemical composition is kept fixed when changing the surface charge density and the supercell dipole $P_z$ becomes a new observable from the simulation. This motivated the development of finite-field DFTMD by exploring
 the constant electric displacement $\bar{D}$ Hamiltonian~\cite{stengel2009electric}. It has been shown that constant $\bar{D}$ simulation significantly speeds up the convergence of the supercell polarization and it also provides a more accurate calculation for the Helmholtz capacitance $C_\mathrm{H}$, as the finite-size error due to the application of periodic boundary conditions is removed~\cite{Zhang:2016ca}. In this setup, the Helmholtz capacitance $C_\mathrm{H}$ can be readily obtained from the supercell polarization $P_z$ using the following expression:
 \begin{equation}
 \label{eq:C_H_constantD}
     C_\mathrm{H}=\frac{1}{2\pi L_z}\left (\frac{\partial \sigma_0}{\partial \langle P_z \rangle} \right )_{\bar{D}} 
 \end{equation}
 where $L_z$ is the supercell length in the direction in which $\bar{D}$ field is applied and $\sigma_0$ is the surface proton-charge density.   

The first study of the protonic double layer using the finite-field DFTMD approach was carried out for the rutile TiO$_{2}$(110)/NaCl electrolyte interface (Fig.~\ref{fig:atomistic_EDL}a). As shown by Zhang et al.~\cite{zhang2019coupling}, the supercell polarization converges quickly by switching the electric boundary condition from constant $\bar{E}$ to constant $\bar{D}$ (Fig.~\ref{fig:atomistic_EDL} b). Moreover, a stronger fluctuation of the water molecules is found at a higher pH as compared to that at a lower pH (Fig.~\ref{fig:atomistic_EDL} c), which leads to a 50\% increase in the Helmholtz capacitance from about 60$\mu$F cm$^{-2}$ to 90$\mu$F cm$^{-2}$. It is also found that What contributes to the Helmholtz capacitance is the proton exchange between the surface group and the solvating water molecules. When both adsorbed water molecules and charged surface groups are constrained to not undergo reaction, $C_\mathrm{H}$ becomes quite close to the previous results~\cite{cheng2014electric}. These findings highlight the importance of the dynamical description in the modelling of the protonic double layer. 

Subsequently, finite-field DFTMD has also been applied to study SnO$_{2}$, which is an isostructure of TiO$_2$ and has a characteristic dissociative water adsorption~\cite{jia2020computing}. Jia et al.~\cite{jia2021origin} computed the Helmholtz capacitance of the SnO$_{2}$(110)/NaCl electrolyte interface and developed a differential Helmholtz capacitance model accordingly. It is found that Dissociative adsorption of water at the surface leads to a higher Helmholtz capacitance (109 $\mu$F cm$^{-2}$) when compared to the molecular adsorption of water (61 $\mu$F cm$^{-2}$). More importantly, it is shown that the dipole at the interface changes the double layer potential significantly and the specific orientation of chemisorbed water molecules leads to an asymmetric profile of the differential Helmholtz capacitance (Fig.~\ref{fig:atomistic_EDL} d). This explains the molecular origin for asymmetric EDLs seen experimentally in a range of metal oxides/hydroxides.

\section{Conclusion and Outlook}

In this chapter, we summarize the fundamentals of the protonic double layer formed at the metal-oxide/electrolyte interface, and discuss the current understanding and the recent progress on this topic from both experimental and theoretical approaches. 

Nevertheless, given the importance of metal-oxide surfaces in practical applications, our understanding of the relationship between the EDL structure and the chemical reactivity is still quite limited. This may not come as a total surprise for an interdisciplinary area where colloid science, electrochemistry and surface science overlap. On one hand, most knowledge of protonic EDL is based on macroscopic observations such as current and concentration from colloid science and electrochemistry, as discussed in this chapter. On the other hand, surface science studies which provided great microscopic insight into various types of surface defects (e.g., oxygen vacancies and surface reconstruction) and surface reactions were mostly performed in ultra-high vacuum (UHV) conditions~\cite{1999.Freund,2015Thorton,Diebold:2017bp}. In this regard, recent advances in experimental techniques such as high-resolution scanning tunnelling microscopy~\cite{Hussain:2016bu}, atomic force microscopy~\cite{Maria13} and apparatus for dosing liquid water in UHV conditions~\cite{Balajka:2018gw} are quite promising to study EDLs at the metal-oxide/electrolyte interface and to bridge this knowledge gap.

Complement to these experimental advances, materials modelling approaches, in particular, finite-field DFTMD simulations have been successfully applied to investigate structure and reactivity of EDLs at the metal-oxide/electrolyte interface, as discussed in this chapter. Further investigations to disentangle the signals from the surface-sensitive vibrational spectroscopy at charged oxide surfaces would be interesting~\cite{Joutsuka:2018bf,2018.Pezzotti}.  Nevertheless, many processes associated with EDLs, e.g. the dissolution of the metal-oxide surface and the charging dynamics, cannot possibly be described with a few hundred atoms and tens of picoseconds typically provided by DFTMD. Therefore, data-driven approaches such as atomistic machine learning could be trans-formative here for modelling the metal-oxide/electrolyte interface~\cite{Shao2021}, as it allows for large-scale simulations with the accuracy of reference electronic structure calculations but only at a fraction of the computational cost~\cite{DeepMD}.


\begin{mcitethebibliography}{67}
\providecommand*\natexlab[1]{#1}
\providecommand*\mciteSetBstSublistMode[1]{}
\providecommand*\mciteSetBstMaxWidthForm[2]{}
\providecommand*\mciteBstWouldAddEndPuncttrue
  {\def\EndOfBibitem{\unskip.}}
\providecommand*\mciteBstWouldAddEndPunctfalse
  {\let\EndOfBibitem\relax}
\providecommand*\mciteSetBstMidEndSepPunct[3]{}
\providecommand*\mciteSetBstSublistLabelBeginEnd[3]{}
\providecommand*\EndOfBibitem{}
\mciteSetBstSublistMode{f}
\mciteSetBstMaxWidthForm{subitem}{(\alph{mcitesubitemcount})}
\mciteSetBstSublistLabelBeginEnd
  {\mcitemaxwidthsubitemform\space}
  {\relax}
  {\relax}

\bibitem[{US Energy Information Administration}(2019)]{US_energy}
{US Energy Information Administration}, Annual Energy Outlook. 2019\relax
\mciteBstWouldAddEndPuncttrue
\mciteSetBstMidEndSepPunct{\mcitedefaultmidpunct}
{\mcitedefaultendpunct}{\mcitedefaultseppunct}\relax
\EndOfBibitem
\bibitem[{OECE}(2011)]{OECD}
{OECE}, OECD Environmental Outlook to 2050. 2011\relax
\mciteBstWouldAddEndPuncttrue
\mciteSetBstMidEndSepPunct{\mcitedefaultmidpunct}
{\mcitedefaultendpunct}{\mcitedefaultseppunct}\relax
\EndOfBibitem
\bibitem[Ardizzone and Trasatti(1996)Ardizzone, and Trasatti]{Ardizzone:1996ca}
Ardizzone,~S.; Trasatti,~S. {Interfacial properties of oxides with
  technological impact in electrochemistry}. \emph{Adv. Colloid Interface Sci.}
  \textbf{1996}, \emph{64}, 173 -- 251\relax
\mciteBstWouldAddEndPuncttrue
\mciteSetBstMidEndSepPunct{\mcitedefaultmidpunct}
{\mcitedefaultendpunct}{\mcitedefaultseppunct}\relax
\EndOfBibitem
\bibitem[Winter and Brodd(2004)Winter, and Brodd]{Winter:2004ge}
Winter,~M.; Brodd,~R.~J. {What are batteries, fuel cells, and supercapacitors?}
  \emph{Chem. Rev.} \textbf{2004}, \emph{104}, 4245 -- 4270\relax
\mciteBstWouldAddEndPuncttrue
\mciteSetBstMidEndSepPunct{\mcitedefaultmidpunct}
{\mcitedefaultendpunct}{\mcitedefaultseppunct}\relax
\EndOfBibitem
\bibitem[Simon and Gogotsi(2008)Simon, and Gogotsi]{Simon:2008dc}
Simon,~P.; Gogotsi,~Y. {Materials for electrochemical capacitors.} \emph{Nat.
  Mater.} \textbf{2008}, \emph{7}, 845 -- 854\relax
\mciteBstWouldAddEndPuncttrue
\mciteSetBstMidEndSepPunct{\mcitedefaultmidpunct}
{\mcitedefaultendpunct}{\mcitedefaultseppunct}\relax
\EndOfBibitem
\bibitem[Song \latin{et~al.}(2018)Song, Bai, Moysiadou, Lee, Hu, Liardet, and
  Hu]{Song.2018}
Song,~F.; Bai,~L.; Moysiadou,~A.; Lee,~S.; Hu,~C.; Liardet,~L.; Hu,~X.
  {Transition metal oxides as electrocatalysts for the oxygen evolution
  reaction in alkaline solutions: An application-inspired renaissance}.
  \emph{J. Am. Chem. Soc.} \textbf{2018}, \emph{140}, 7748--7759\relax
\mciteBstWouldAddEndPuncttrue
\mciteSetBstMidEndSepPunct{\mcitedefaultmidpunct}
{\mcitedefaultendpunct}{\mcitedefaultseppunct}\relax
\EndOfBibitem
\bibitem[Ren \latin{et~al.}(2012)Ren, Ma, and Bruce]{Ren:2012go}
Ren,~Y.; Ma,~Z.; Bruce,~P.~G. {Ordered mesoporous metal oxides: Synthesis and
  applications}. \emph{Chem. Soc. Rev.} \textbf{2012}, \emph{41}\relax
\mciteBstWouldAddEndPuncttrue
\mciteSetBstMidEndSepPunct{\mcitedefaultmidpunct}
{\mcitedefaultendpunct}{\mcitedefaultseppunct}\relax
\EndOfBibitem
\bibitem[Zheng \latin{et~al.}(2018)Zheng, Tang, Li, Hu, Zhang, Xue, and
  Pang]{Zheng:2018ie}
Zheng,~M.; Tang,~H.; Li,~L.; Hu,~Q.; Zhang,~L.; Xue,~H.; Pang,~H.
  {Hierarchically nanostructured transition metal oxides for lithium-ion
  batteries}. \emph{Adv. Sci.} \textbf{2018}, \emph{5}, 1700592 -- 24\relax
\mciteBstWouldAddEndPuncttrue
\mciteSetBstMidEndSepPunct{\mcitedefaultmidpunct}
{\mcitedefaultendpunct}{\mcitedefaultseppunct}\relax
\EndOfBibitem
\bibitem[Raghunath and Perumal(2017)Raghunath, and Perumal]{Raghunath:2017hk}
Raghunath,~A.; Perumal,~E. {Metal oxide nanoparticles as antimicrobial agents:
  A promise for the future}. \emph{Int. J. Antimicrob. Agents} \textbf{2017},
  \emph{49}, 137 -- 152\relax
\mciteBstWouldAddEndPuncttrue
\mciteSetBstMidEndSepPunct{\mcitedefaultmidpunct}
{\mcitedefaultendpunct}{\mcitedefaultseppunct}\relax
\EndOfBibitem
\bibitem[Lokhande \latin{et~al.}(2011)Lokhande, Dubal, and Joo]{Lokhande11}
Lokhande,~C.; Dubal,~D.; Joo,~O.-S. Metal oxide thin film based
  supercapacitors. \emph{Curr. Appl. Phys.} \textbf{2011}, \emph{11},
  255--270\relax
\mciteBstWouldAddEndPuncttrue
\mciteSetBstMidEndSepPunct{\mcitedefaultmidpunct}
{\mcitedefaultendpunct}{\mcitedefaultseppunct}\relax
\EndOfBibitem
\bibitem[Joo and Zhao(2017)Joo, and Zhao]{JOO201729}
Joo,~S.~H.; Zhao,~D. Environmental dynamics of metal oxide nanoparticles in
  heterogeneous systems: A review. \emph{J. Hazard. Mater.} \textbf{2017},
  \emph{322}, 29--47\relax
\mciteBstWouldAddEndPuncttrue
\mciteSetBstMidEndSepPunct{\mcitedefaultmidpunct}
{\mcitedefaultendpunct}{\mcitedefaultseppunct}\relax
\EndOfBibitem
\bibitem[Mu \latin{et~al.}(2017)Mu, Zhao, Dohnálek, and Gong]{2017.Mu}
Mu,~R.; Zhao,~Z.-j.; Dohnálek,~Z.; Gong,~J. {Structural motifs of water on
  metal oxide surfaces}. \emph{Chem. Soc. Rev.} \textbf{2017}, \emph{46},
  1785--1806\relax
\mciteBstWouldAddEndPuncttrue
\mciteSetBstMidEndSepPunct{\mcitedefaultmidpunct}
{\mcitedefaultendpunct}{\mcitedefaultseppunct}\relax
\EndOfBibitem
\bibitem[Lyons \latin{et~al.}(2017)Lyons, Doyle, Browne, Godwin, and
  Rovetta]{Lyons:2017cg}
Lyons,~M. E.~G.; Doyle,~R.~L.; Browne,~M.~P.; Godwin,~I.~J.; Rovetta,~A. A.~S.
  {Recent developments in electrochemical water oxidation}. \emph{Curr. Opin.
  Electrochem.} \textbf{2017}, \emph{1}, 40 -- 45\relax
\mciteBstWouldAddEndPuncttrue
\mciteSetBstMidEndSepPunct{\mcitedefaultmidpunct}
{\mcitedefaultendpunct}{\mcitedefaultseppunct}\relax
\EndOfBibitem
\bibitem[Sato(1998)]{Sato1998}
Sato,~N. \emph{{Electrochemistry at metal and semiconductor electrodes}};
  Elsevier: Amsterdam, 1998\relax
\mciteBstWouldAddEndPuncttrue
\mciteSetBstMidEndSepPunct{\mcitedefaultmidpunct}
{\mcitedefaultendpunct}{\mcitedefaultseppunct}\relax
\EndOfBibitem
\bibitem[Br\o{}nsted(1923)]{Bronsted23}
Br\o{}nsted,~J.~N. Einige Bemerkungen über den Begriff der Säuren und Basen.
  \emph{Recl. Trav. Chim. Pays-Bas} \textbf{1923}, \emph{42}, 718--728\relax
\mciteBstWouldAddEndPuncttrue
\mciteSetBstMidEndSepPunct{\mcitedefaultmidpunct}
{\mcitedefaultendpunct}{\mcitedefaultseppunct}\relax
\EndOfBibitem
\bibitem[Lowry(1923)]{Lowry23}
Lowry,~T.~M. The uniqueness of hydrogen. \emph{J. Soc. Chem. Ind.}
  \textbf{1923}, \emph{42}, 43--47\relax
\mciteBstWouldAddEndPuncttrue
\mciteSetBstMidEndSepPunct{\mcitedefaultmidpunct}
{\mcitedefaultendpunct}{\mcitedefaultseppunct}\relax
\EndOfBibitem
\bibitem[Cheng \latin{et~al.}(2014)Cheng, Liu, VandeVondele, Sulpizi, and
  Sprik]{Cheng:2014jb}
Cheng,~J.; Liu,~X.; VandeVondele,~J.; Sulpizi,~M.; Sprik,~M. {Redox potentials
  and acidity constants from density functional theory based molecular
  dynamics}. \emph{Acc. Chem. Res.} \textbf{2014}, \emph{47}, 3522 --
  3529\relax
\mciteBstWouldAddEndPuncttrue
\mciteSetBstMidEndSepPunct{\mcitedefaultmidpunct}
{\mcitedefaultendpunct}{\mcitedefaultseppunct}\relax
\EndOfBibitem
\bibitem[Noguera(1996)]{noguera1996physics}
Noguera,~C. \emph{Physics and chemistry at oxide surfaces}; Cambridge
  University Press, 1996\relax
\mciteBstWouldAddEndPuncttrue
\mciteSetBstMidEndSepPunct{\mcitedefaultmidpunct}
{\mcitedefaultendpunct}{\mcitedefaultseppunct}\relax
\EndOfBibitem
\bibitem[Lyklema(2005)]{lyklema2005fundamentals}
Lyklema,~J. \emph{Fundamentals of interface and colloid science: Soft
  colloids}; Elsevier, 2005; Vol.~5\relax
\mciteBstWouldAddEndPuncttrue
\mciteSetBstMidEndSepPunct{\mcitedefaultmidpunct}
{\mcitedefaultendpunct}{\mcitedefaultseppunct}\relax
\EndOfBibitem
\bibitem[Schmickler and Santos(2010)Schmickler, and
  Santos]{schmickler2010interfacial}
Schmickler,~W.; Santos,~E. \emph{Interfacial electrochemistry}; Springer
  Science \& Business Media, 2010\relax
\mciteBstWouldAddEndPuncttrue
\mciteSetBstMidEndSepPunct{\mcitedefaultmidpunct}
{\mcitedefaultendpunct}{\mcitedefaultseppunct}\relax
\EndOfBibitem
\bibitem[Bunker and Casey(2020)Bunker, and Casey]{bunker_aqueous_2020}
Bunker,~B.~C.; Casey,~W.~H. \emph{The aqueous chemistry of oxides}; Oxford
  University Press, 2020\relax
\mciteBstWouldAddEndPuncttrue
\mciteSetBstMidEndSepPunct{\mcitedefaultmidpunct}
{\mcitedefaultendpunct}{\mcitedefaultseppunct}\relax
\EndOfBibitem
\bibitem[Memming(2015)]{memming2015semiconductor}
Memming,~R. \emph{Semiconductor electrochemistry}; John Wiley \& Sons,
  2015\relax
\mciteBstWouldAddEndPuncttrue
\mciteSetBstMidEndSepPunct{\mcitedefaultmidpunct}
{\mcitedefaultendpunct}{\mcitedefaultseppunct}\relax
\EndOfBibitem
\bibitem[Gr\"atzel(2001)]{2001.Graetzel}
Gr\"atzel,~M. {Photoelectrochemical cells}. \emph{Nature} \textbf{2001},
  \emph{414}, 338--344\relax
\mciteBstWouldAddEndPuncttrue
\mciteSetBstMidEndSepPunct{\mcitedefaultmidpunct}
{\mcitedefaultendpunct}{\mcitedefaultseppunct}\relax
\EndOfBibitem
\bibitem[Lützenkirchen \latin{et~al.}(2012)Lützenkirchen, Preočanin,
  Kovačević, Tomišić, Lövgren, and Kallay]{2012luetzenkirchen}
Lützenkirchen,~J.; Preočanin,~T.; Kovačević,~D.; Tomišić,~V.;
  Lövgren,~L.; Kallay,~N. {Potentiometric titrations as a tool for surface
  charge determination}. \emph{Croat. Chem. Acta} \textbf{2012}, \emph{85}, 391
  -- 417\relax
\mciteBstWouldAddEndPuncttrue
\mciteSetBstMidEndSepPunct{\mcitedefaultmidpunct}
{\mcitedefaultendpunct}{\mcitedefaultseppunct}\relax
\EndOfBibitem
\bibitem[B{\'e}rub{\'e} and De~Bruyn(1968)B{\'e}rub{\'e}, and
  De~Bruyn]{Berube:1968im}
B{\'e}rub{\'e},~Y.~G.; De~Bruyn,~P.~L. {Adsorption at the rutile-solution
  interface. II. Model of the electrochemical double layer}. \emph{J. Colloid
  Interface Sci.} \textbf{1968}, \emph{28}, 92--105\relax
\mciteBstWouldAddEndPuncttrue
\mciteSetBstMidEndSepPunct{\mcitedefaultmidpunct}
{\mcitedefaultendpunct}{\mcitedefaultseppunct}\relax
\EndOfBibitem
\bibitem[Lvovich(2012)]{lvovich_impedance_2015}
Lvovich,~V.~F. \emph{Impedance {spectroscopy}: {Applications} to
  {electrochemical} and {dielectric} {phenomena}}; Wiley, 2012\relax
\mciteBstWouldAddEndPuncttrue
\mciteSetBstMidEndSepPunct{\mcitedefaultmidpunct}
{\mcitedefaultendpunct}{\mcitedefaultseppunct}\relax
\EndOfBibitem
\bibitem[Tomkiewicz(1979)]{Tomkiewicz:1979ifa}
Tomkiewicz,~M. {The potential distribution at the TiO$_2$ aqueous electrolyte
  interface}. \emph{J. Electrochem. Soc.} \textbf{1979}, \emph{126}, 1505 --
  1510\relax
\mciteBstWouldAddEndPuncttrue
\mciteSetBstMidEndSepPunct{\mcitedefaultmidpunct}
{\mcitedefaultendpunct}{\mcitedefaultseppunct}\relax
\EndOfBibitem
\bibitem[Faulkner and Bard(2002)Faulkner, and
  Bard]{faulkner2002electrochemical}
Faulkner,~L.~R.; Bard,~A.~J. \emph{Electrochemical methods: Fundamentals and
  applications}; John Wiley and Sons, 2002\relax
\mciteBstWouldAddEndPuncttrue
\mciteSetBstMidEndSepPunct{\mcitedefaultmidpunct}
{\mcitedefaultendpunct}{\mcitedefaultseppunct}\relax
\EndOfBibitem
\bibitem[Sugimoto \latin{et~al.}(2006)Sugimoto, Yokoshima, Murakami, and
  Takasu]{2006.Sugimoto}
Sugimoto,~W.; Yokoshima,~K.; Murakami,~Y.; Takasu,~Y. {Charge storage mechanism
  of nanostructured anhydrous and hydrous ruthenium-based oxides}.
  \emph{Electrochim. Acta} \textbf{2006}, \emph{52}, 1742--1748\relax
\mciteBstWouldAddEndPuncttrue
\mciteSetBstMidEndSepPunct{\mcitedefaultmidpunct}
{\mcitedefaultendpunct}{\mcitedefaultseppunct}\relax
\EndOfBibitem
\bibitem[Ong \latin{et~al.}(1992)Ong, Zhao, and Eisenthal]{Ong:1992ca}
Ong,~S.; Zhao,~X.; Eisenthal,~K.~B. {Polarization of water molecules at a
  charged interface: Second harmonic studies of the silica/water\ interface}.
  \emph{Chem. Phys. Lett.} \textbf{1992}, \emph{191}, 327--335\relax
\mciteBstWouldAddEndPuncttrue
\mciteSetBstMidEndSepPunct{\mcitedefaultmidpunct}
{\mcitedefaultendpunct}{\mcitedefaultseppunct}\relax
\EndOfBibitem
\bibitem[Bischoff \latin{et~al.}(2020)Bischoff, Biriukov, Předota, Roke, and
  Marchioro]{2020.Bischoff}
Bischoff,~M.; Biriukov,~D.; Předota,~M.; Roke,~S.; Marchioro,~A. {Surface
  potential and interfacial water order at the amorphous {T}i{O}$_2$
  nanoparticle/aqueous interface}. \emph{J. Phys. Chem. C} \textbf{2020},
  \emph{124}, 10961--10974\relax
\mciteBstWouldAddEndPuncttrue
\mciteSetBstMidEndSepPunct{\mcitedefaultmidpunct}
{\mcitedefaultendpunct}{\mcitedefaultseppunct}\relax
\EndOfBibitem
\bibitem[Brown \latin{et~al.}(2016)Brown, Abbas, Kleibert, Green, Goel, May,
  and Squires]{Brown:2016bo}
Brown,~M.~A.; Abbas,~Z.; Kleibert,~A.; Green,~R.~G.; Goel,~A.; May,~.~S.;
  Squires,~T.~M. {Determination of surface potential and electrical
  double-layer structure at the aqueous electrolyte-nanoparticle interface}.
  \emph{Phys. Rev. X} \textbf{2016}, \emph{6}, 011007 -- 12\relax
\mciteBstWouldAddEndPuncttrue
\mciteSetBstMidEndSepPunct{\mcitedefaultmidpunct}
{\mcitedefaultendpunct}{\mcitedefaultseppunct}\relax
\EndOfBibitem
\bibitem[Brown \latin{et~al.}(2016)Brown, Goel, and Abbas]{Brown:2016kwa}
Brown,~M.~A.; Goel,~A.; Abbas,~Z. {Effect of electrolyte concentration on the
  Stern layer thickness at a charged interface}. \emph{Angew. Chem. Int. Ed.}
  \textbf{2016}, \emph{128}, 3854 -- 3858\relax
\mciteBstWouldAddEndPuncttrue
\mciteSetBstMidEndSepPunct{\mcitedefaultmidpunct}
{\mcitedefaultendpunct}{\mcitedefaultseppunct}\relax
\EndOfBibitem
\bibitem[Parks(1965)]{1965.Parks}
Parks,~G.~A. {The isoelectric points of solid oxides, solid hydroxides, and
  aqueous hydroxo complex systems}. \emph{Chem. Rev.} \textbf{1965}, \emph{65},
  177--198\relax
\mciteBstWouldAddEndPuncttrue
\mciteSetBstMidEndSepPunct{\mcitedefaultmidpunct}
{\mcitedefaultendpunct}{\mcitedefaultseppunct}\relax
\EndOfBibitem
\bibitem[Schindler and Kamber(1968)Schindler, and Kamber]{1968.Schindler}
Schindler,~P.; Kamber,~H.~R. {Die Acidität von Silanolgruppen. Vorläufige
  Mitteillung}. \emph{Helv. Chim. Acta} \textbf{1968}, \emph{51},
  1781--1786\relax
\mciteBstWouldAddEndPuncttrue
\mciteSetBstMidEndSepPunct{\mcitedefaultmidpunct}
{\mcitedefaultendpunct}{\mcitedefaultseppunct}\relax
\EndOfBibitem
\bibitem[Stumm \latin{et~al.}(1970)Stumm, Huang, and Jenkins]{Stumm70}
Stumm,~W.; Huang,~C.~P.; Jenkins,~S.~R. {Specific chemical interaction
  affecting the stability of dispersed systems}. \emph{Croat. Chem. Acta}
  \textbf{1970}, \emph{42}, 223 -- 244\relax
\mciteBstWouldAddEndPuncttrue
\mciteSetBstMidEndSepPunct{\mcitedefaultmidpunct}
{\mcitedefaultendpunct}{\mcitedefaultseppunct}\relax
\EndOfBibitem
\bibitem[Davis \latin{et~al.}(1978)Davis, James, and Leckie]{DAVIS:1978fn}
Davis,~J.~A.; James,~R.~O.; Leckie,~J.~O. {Surface ionization and complexation
  at the oxide/water interface: I. Computation of electrical double layer
  properties in simple elecrolytes}. \emph{J. Colloid Interface Sci.}
  \textbf{1978}, \emph{63}, 480 -- 499\relax
\mciteBstWouldAddEndPuncttrue
\mciteSetBstMidEndSepPunct{\mcitedefaultmidpunct}
{\mcitedefaultendpunct}{\mcitedefaultseppunct}\relax
\EndOfBibitem
\bibitem[Westall and Hohl(1980)Westall, and Hohl]{1980.Westall}
Westall,~J.; Hohl,~H. {A comparison of electrostatic models for the
  oxide/solution interface}. \emph{Adv. Colloid Interface Sci.} \textbf{1980},
  \emph{12}, 265--294\relax
\mciteBstWouldAddEndPuncttrue
\mciteSetBstMidEndSepPunct{\mcitedefaultmidpunct}
{\mcitedefaultendpunct}{\mcitedefaultseppunct}\relax
\EndOfBibitem
\bibitem[Hiemstra \latin{et~al.}(1989)Hiemstra, Riemsdijk, and
  Bolt]{Hiemstra:1989vh}
Hiemstra,~T.; Riemsdijk,~W.~v.; Bolt,~G. {Multisite proton adsorption modeling
  at the solid/solution interface of (hydr)oxides: A new approach I. Model
  description and evaluat\ ion of intrinsic reaction constants}. \emph{J.
  Colloid Interface Sci} \textbf{1989}, \emph{133}, 91--104\relax
\mciteBstWouldAddEndPuncttrue
\mciteSetBstMidEndSepPunct{\mcitedefaultmidpunct}
{\mcitedefaultendpunct}{\mcitedefaultseppunct}\relax
\EndOfBibitem
\bibitem[Hiemstra and van Riemsdijk(1996)Hiemstra, and van
  Riemsdijk]{hiemstra1996surface}
Hiemstra,~T.; van Riemsdijk,~W.~H. A surface structural approach to ion
  adsorption: The charge distribution {(CD)} model. \emph{Journal of colloid
  and interface science} \textbf{1996}, \emph{179}, 488--508\relax
\mciteBstWouldAddEndPuncttrue
\mciteSetBstMidEndSepPunct{\mcitedefaultmidpunct}
{\mcitedefaultendpunct}{\mcitedefaultseppunct}\relax
\EndOfBibitem
\bibitem[Brown~Jr and Parks(2001)Brown~Jr, and Parks]{2001brown}
Brown~Jr,~G.~E.; Parks,~G.~A. {Sorption of trace elements on mineral surfaces:
  Modern perspectives from spectroscopic studies, and comments on sorption in
  the marine environment}. \emph{Int. Geol. Rev.} \textbf{2001}, \emph{43},
  963--1073\relax
\mciteBstWouldAddEndPuncttrue
\mciteSetBstMidEndSepPunct{\mcitedefaultmidpunct}
{\mcitedefaultendpunct}{\mcitedefaultseppunct}\relax
\EndOfBibitem
\bibitem[Bickmore \latin{et~al.}(2006)Bickmore, Rosso, and
  Mitchell]{2006.Bickmore}
Bickmore,~B.; Rosso,~K.; Mitchell,~S. In \emph{Surface Complexation Modelling};
  Lützenkirchen,~J., Ed.; Elsevier, 2006; Chapter 9, pp 269--283\relax
\mciteBstWouldAddEndPuncttrue
\mciteSetBstMidEndSepPunct{\mcitedefaultmidpunct}
{\mcitedefaultendpunct}{\mcitedefaultseppunct}\relax
\EndOfBibitem
\bibitem[Car and Parrinello(1985)Car, and Parrinello]{Car:1985ix}
Car,~R.; Parrinello,~M. {Unified approach for molecular dynamics and
  density-functional theory.} \emph{Phys. Rev. Lett.} \textbf{1985}, \emph{55},
  2471 -- 2474\relax
\mciteBstWouldAddEndPuncttrue
\mciteSetBstMidEndSepPunct{\mcitedefaultmidpunct}
{\mcitedefaultendpunct}{\mcitedefaultseppunct}\relax
\EndOfBibitem
\bibitem[Marx and Hutter(2012)Marx, and Hutter]{marx09}
Marx,~D.; Hutter,~J. \emph{{Ab initio molecular dynamics: Basic theory and
  advanced methods}}; Cambridge University Press, 2012\relax
\mciteBstWouldAddEndPuncttrue
\mciteSetBstMidEndSepPunct{\mcitedefaultmidpunct}
{\mcitedefaultendpunct}{\mcitedefaultseppunct}\relax
\EndOfBibitem
\bibitem[Předota \latin{et~al.}(2004)Předota, Bandura, Cummings, Kubicki,
  Wesolowski, Chialvo, and Machesky]{Predota:2004cy}
Předota,~M.; Bandura,~A.~V.; Cummings,~P.~T.; Kubicki,~J.~D.;
  Wesolowski,~D.~J.; Chialvo,~A.~A.; Machesky,~M.~L. {Electric double layer at
  the rutile (110) surface. 1. Structure of surfaces and interfacial water from
  molecular dynamics by use of ab initio potentials}. \emph{J. Phys. Chem. B}
  \textbf{2004}, \emph{108}, 12049 -- 12060\relax
\mciteBstWouldAddEndPuncttrue
\mciteSetBstMidEndSepPunct{\mcitedefaultmidpunct}
{\mcitedefaultendpunct}{\mcitedefaultseppunct}\relax
\EndOfBibitem
\bibitem[Parez \latin{et~al.}(2014)Parez, Předota, and Machesky]{Parez:2014dx}
Parez,~S.; Předota,~M.; Machesky,~M. {Dielectric properties of water at rutile
  and graphite surfaces: Effect of molecular structure}. \emph{J. Phys. Chem.
  C} \textbf{2014}, \emph{118}, 4818 -- 4834\relax
\mciteBstWouldAddEndPuncttrue
\mciteSetBstMidEndSepPunct{\mcitedefaultmidpunct}
{\mcitedefaultendpunct}{\mcitedefaultseppunct}\relax
\EndOfBibitem
\bibitem[Bandura and Kubicki(2003)Bandura, and Kubicki]{Bandura:2003fq}
Bandura,~A.~V.; Kubicki,~J.~D. {Derivation of force field parameters for
  {T}i{O}$_2$-{H}$_2${O} systems from ab initio calculations}. \emph{J. Phys.
  Chem. B} \textbf{2003}, \emph{107}, 11072 -- 11081\relax
\mciteBstWouldAddEndPuncttrue
\mciteSetBstMidEndSepPunct{\mcitedefaultmidpunct}
{\mcitedefaultendpunct}{\mcitedefaultseppunct}\relax
\EndOfBibitem
\bibitem[Cygan \latin{et~al.}(2004)Cygan, Liang, and
  Kalinichev]{cygan2004molecular}
Cygan,~R.~T.; Liang,~J.-J.; Kalinichev,~A.~G. Molecular models of hydroxide,
  oxyhydroxide, and clay phases and the development of a general force field.
  \emph{J. Phys. Chem. B} \textbf{2004}, \emph{108}, 1255--1266\relax
\mciteBstWouldAddEndPuncttrue
\mciteSetBstMidEndSepPunct{\mcitedefaultmidpunct}
{\mcitedefaultendpunct}{\mcitedefaultseppunct}\relax
\EndOfBibitem
\bibitem[Heinz \latin{et~al.}(2013)Heinz, Lin, Mishra, and Emami]{2013.Heinz}
Heinz,~H.; Lin,~T.-J.; Mishra,~R.~K.; Emami,~F.~S. {Thermodynamically
  consistent force fields for the assembly of inorganic, organic, and
  niological nanostructures: The INTERFACE force field}. \emph{Langmuir}
  \textbf{2013}, \emph{29}, 1754--1765\relax
\mciteBstWouldAddEndPuncttrue
\mciteSetBstMidEndSepPunct{\mcitedefaultmidpunct}
{\mcitedefaultendpunct}{\mcitedefaultseppunct}\relax
\EndOfBibitem
\bibitem[Cheng and Sprik(2014)Cheng, and Sprik]{cheng2014electric}
Cheng,~J.; Sprik,~M. The electric double layer at a rutile {T}i{O}$_2$ water
  interface modelled using density functional theory based molecular dynamics
  simulation. \emph{J. Phys: Condens. Matter} \textbf{2014}, \emph{26},
  244108\relax
\mciteBstWouldAddEndPuncttrue
\mciteSetBstMidEndSepPunct{\mcitedefaultmidpunct}
{\mcitedefaultendpunct}{\mcitedefaultseppunct}\relax
\EndOfBibitem
\bibitem[Ulman \latin{et~al.}(2019)Ulman, Poli, Seriani, Piccinin, and
  Gebauer]{ulman2019understanding}
Ulman,~K.; Poli,~E.; Seriani,~N.; Piccinin,~S.; Gebauer,~R. Understanding the
  electrochemical double layer at the hematite/water interface: A first
  principles molecular dynamics study. \emph{J. Chem. Phys.} \textbf{2019},
  \emph{150}, 041707\relax
\mciteBstWouldAddEndPuncttrue
\mciteSetBstMidEndSepPunct{\mcitedefaultmidpunct}
{\mcitedefaultendpunct}{\mcitedefaultseppunct}\relax
\EndOfBibitem
\bibitem[Zhang \latin{et~al.}(2019)Zhang, Hutter, and Sprik]{zhang2019coupling}
Zhang,~C.; Hutter,~J.; Sprik,~M. Coupling of surface chemistry and electric
  double layer at {T}i{O}$_2$ electrochemical interfaces. \emph{J. Phys. Chem.
  Lett.} \textbf{2019}, \emph{10}, 3871--3876\relax
\mciteBstWouldAddEndPuncttrue
\mciteSetBstMidEndSepPunct{\mcitedefaultmidpunct}
{\mcitedefaultendpunct}{\mcitedefaultseppunct}\relax
\EndOfBibitem
\bibitem[Jia \latin{et~al.}(2021)Jia, Zhang, and Cheng]{jia2021origin}
Jia,~M.; Zhang,~C.; Cheng,~J. Origin of Asymmetric Electric Double Layers at
  Electrified Oxide/Electrolyte Interfaces. \emph{J. Phys. Chem. Lett.}
  \textbf{2021}, \emph{12}, 4616--4622\relax
\mciteBstWouldAddEndPuncttrue
\mciteSetBstMidEndSepPunct{\mcitedefaultmidpunct}
{\mcitedefaultendpunct}{\mcitedefaultseppunct}\relax
\EndOfBibitem
\bibitem[Zhang \latin{et~al.}(2020)Zhang, Sayer, Hutter, and
  Sprik]{zhang2020modelling}
Zhang,~C.; Sayer,~T.; Hutter,~J.; Sprik,~M. Modelling electrochemical systems
  with finite field molecular dynamics. \emph{J. Phys: Energy} \textbf{2020},
  \emph{2}, 032005\relax
\mciteBstWouldAddEndPuncttrue
\mciteSetBstMidEndSepPunct{\mcitedefaultmidpunct}
{\mcitedefaultendpunct}{\mcitedefaultseppunct}\relax
\EndOfBibitem
\bibitem[Stengel \latin{et~al.}(2009)Stengel, Spaldin, and
  Vanderbilt]{stengel2009electric}
Stengel,~M.; Spaldin,~N.~A.; Vanderbilt,~D. Electric displacement as the
  fundamental variable in electronic-structure calculations. \emph{Nat. Phys.}
  \textbf{2009}, \emph{5}, 304--308\relax
\mciteBstWouldAddEndPuncttrue
\mciteSetBstMidEndSepPunct{\mcitedefaultmidpunct}
{\mcitedefaultendpunct}{\mcitedefaultseppunct}\relax
\EndOfBibitem
\bibitem[Zhang and Sprik(2016)Zhang, and Sprik]{Zhang:2016ca}
Zhang,~C.; Sprik,~M. {Finite field methods for the supercell modeling of
  charged insulator/electrolyte interfaces}. \emph{Phys. Rev. B} \textbf{2016},
  \emph{94}, 245309\relax
\mciteBstWouldAddEndPuncttrue
\mciteSetBstMidEndSepPunct{\mcitedefaultmidpunct}
{\mcitedefaultendpunct}{\mcitedefaultseppunct}\relax
\EndOfBibitem
\bibitem[Jia \latin{et~al.}(2020)Jia, Zhang, Cox, Sprik, and
  Cheng]{jia2020computing}
Jia,~M.; Zhang,~C.; Cox,~S.~J.; Sprik,~M.; Cheng,~J. Computing surface acidity
  constants of proton hopping groups from density functional theory-based
  molecular dynamics: Application to the {S}n{O}$_2$ (110)/{H}$_2${O}
  interface. \emph{J. Chem. Theory Comput.} \textbf{2020}, \emph{16},
  6520--6527\relax
\mciteBstWouldAddEndPuncttrue
\mciteSetBstMidEndSepPunct{\mcitedefaultmidpunct}
{\mcitedefaultendpunct}{\mcitedefaultseppunct}\relax
\EndOfBibitem
\bibitem[Freund(1999)]{1999.Freund}
Freund,~H.-J. {Introductory lecture: Oxide surfaces}. \emph{Faraday Disc.}
  \textbf{1999}, \emph{114}, 1--31\relax
\mciteBstWouldAddEndPuncttrue
\mciteSetBstMidEndSepPunct{\mcitedefaultmidpunct}
{\mcitedefaultendpunct}{\mcitedefaultseppunct}\relax
\EndOfBibitem
\bibitem[Jupille and Thornton(2015)Jupille, and Thornton]{2015Thorton}
Jupille,~J.; Thornton,~G. \emph{{Defects at oxide surfaces}}; Springer,
  2015\relax
\mciteBstWouldAddEndPuncttrue
\mciteSetBstMidEndSepPunct{\mcitedefaultmidpunct}
{\mcitedefaultendpunct}{\mcitedefaultseppunct}\relax
\EndOfBibitem
\bibitem[Diebold(2017)]{Diebold:2017bp}
Diebold,~U. {Perspective: A controversial benchmark system for water-oxide
  interfaces: H$_2$O/TiO$_2$(110)}. \emph{J. Chem. Phys.} \textbf{2017},
  \emph{147}, 040901 -- 4\relax
\mciteBstWouldAddEndPuncttrue
\mciteSetBstMidEndSepPunct{\mcitedefaultmidpunct}
{\mcitedefaultendpunct}{\mcitedefaultseppunct}\relax
\EndOfBibitem
\bibitem[Hussain \latin{et~al.}(2016)Hussain, Tocci, Woolcot, Torrelles, Pang,
  Humphrey, Yim, Grinter, Cabailh, Bikondoa, L\~indsay, Zegenhagen,
  Michaelides, and Thornton]{Hussain:2016bu}
Hussain,~H.; Tocci,~G.; Woolcot,~T.; Torrelles,~X.; Pang,~C.~L.;
  Humphrey,~D.~S.; Yim,~C.~M.; Grinter,~D.~C.; Cabailh,~G.; Bikondoa,~O.;
  L\~indsay,~R.; Zegenhagen,~J.; Michaelides,~A.; Thornton,~G. {Structure of a
  model TiO$_2$ photocatalytic interface}. \emph{Nat. Mater.} \textbf{2016}, 1
  -- 7\relax
\mciteBstWouldAddEndPuncttrue
\mciteSetBstMidEndSepPunct{\mcitedefaultmidpunct}
{\mcitedefaultendpunct}{\mcitedefaultseppunct}\relax
\EndOfBibitem
\bibitem[Ricci \latin{et~al.}(2013)Ricci, Spijker, Stellacci, Molinari, and
  Voïtchovsky]{Maria13}
Ricci,~M.; Spijker,~P.; Stellacci,~F.; Molinari,~J.-F.; Voïtchovsky,~K.
  {Direct visualization of single ions in the Stern layer of calcite}.
  \emph{Langmuir} \textbf{2013}, \emph{29}, 2207--2216\relax
\mciteBstWouldAddEndPuncttrue
\mciteSetBstMidEndSepPunct{\mcitedefaultmidpunct}
{\mcitedefaultendpunct}{\mcitedefaultseppunct}\relax
\EndOfBibitem
\bibitem[Balajka \latin{et~al.}(2018)Balajka, Pavelec, Komora, Schmid, and
  Diebold]{Balajka:2018gw}
Balajka,~J.; Pavelec,~J.; Komora,~M.; Schmid,~M.; Diebold,~U. {Apparatus for
  dosing liquid water in ultrahigh vacuum}. \emph{Rev. Sci. Instrum.}
  \textbf{2018}, \emph{89}, 083906 -- 7\relax
\mciteBstWouldAddEndPuncttrue
\mciteSetBstMidEndSepPunct{\mcitedefaultmidpunct}
{\mcitedefaultendpunct}{\mcitedefaultseppunct}\relax
\EndOfBibitem
\bibitem[Joutsuka \latin{et~al.}(2018)Joutsuka, Hirano, Sprik, and
  Morita]{Joutsuka:2018bf}
Joutsuka,~T.; Hirano,~T.; Sprik,~M.; Morita,~A. {Effects of third-order
  susceptibility in sum frequency generation spectra: a molecular dynamics\
  study in liquid water}. \emph{Phys. Chem. Chem. Phys.} \textbf{2018},
  \emph{20}, 3040 -- 3053\relax
\mciteBstWouldAddEndPuncttrue
\mciteSetBstMidEndSepPunct{\mcitedefaultmidpunct}
{\mcitedefaultendpunct}{\mcitedefaultseppunct}\relax
\EndOfBibitem
\bibitem[Pezzotti \latin{et~al.}(2018)Pezzotti, Galimberti, Shen, and
  Gaigeot]{2018.Pezzotti}
Pezzotti,~S.; Galimberti,~D.~R.; Shen,~Y.~R.; Gaigeot,~M.-P. {Structural
  definition of the BIL and DL: a new universal methodology to rationalize
  non-linear $\chi^{(2)}(\omega)$ SFG signals at charged interfaces, including
  $\chi^{(3)}(\omega)$ contributions}. \emph{Phys. Chem. Chem. Phys.}
  \textbf{2018}, \emph{20}, 5190--5199\relax
\mciteBstWouldAddEndPuncttrue
\mciteSetBstMidEndSepPunct{\mcitedefaultmidpunct}
{\mcitedefaultendpunct}{\mcitedefaultseppunct}\relax
\EndOfBibitem
\bibitem[Shao \latin{et~al.}(2021)Shao, Knijff, Dietrich, Hermansson, and
  Zhang]{Shao2021}
Shao,~Y.; Knijff,~L.; Dietrich,~F.~M.; Hermansson,~K.; Zhang,~C. {Modelling
  bulk electrolytes and electrolyte interfaces with atomistic machine
  learning}. \emph{Batteries \& Supercaps} \textbf{2021}, \emph{4}, 585 --
  595\relax
\mciteBstWouldAddEndPuncttrue
\mciteSetBstMidEndSepPunct{\mcitedefaultmidpunct}
{\mcitedefaultendpunct}{\mcitedefaultseppunct}\relax
\EndOfBibitem
\bibitem[Jia \latin{et~al.}(2020)Jia, Wang, Chen, Lu, Lin, Car, Weinan, and
  Zhang]{DeepMD}
Jia,~W.; Wang,~H.; Chen,~M.; Lu,~D.; Lin,~L.; Car,~R.; Weinan,~E.; Zhang,~L.
  Pushing the Limit of Molecular Dynamics with Ab initio Accuracy to 100
  Million Atoms with Machine Learning. SC20: International Conference for High
  Performance Computing, Networking, Storage and Analysis. 2020; pp 1--14\relax
\mciteBstWouldAddEndPuncttrue
\mciteSetBstMidEndSepPunct{\mcitedefaultmidpunct}
{\mcitedefaultendpunct}{\mcitedefaultseppunct}\relax
\EndOfBibitem
\end{mcitethebibliography}

\providecommand{\latin}[1]{#1}
\makeatletter
\providecommand{\doi}
  {\begingroup\let\do\@makeother\dospecials
  \catcode`\{=1 \catcode`\}=2 \doi@aux}
\providecommand{\doi@aux}[1]{\endgroup\texttt{#1}}
\makeatother
\providecommand*\mcitethebibliography{\thebibliography}
\csname @ifundefined\endcsname{endmcitethebibliography}
  {\let\endmcitethebibliography\endthebibliography}{}






\end{document}